\documentclass{article}
\usepackage{spconf,amsmath,graphicx}

\usepackage{cite}
\usepackage{algorithmic}
\usepackage{array}
\usepackage{stfloats}
\usepackage[export]{adjustbox}
\usepackage{systeme}
\usepackage[caption=false,font=footnotesize]{subfig}

\makeatletter
\def\thickhline{%
  \noalign{\ifnum0=`}\fi\hrule \@height \thickarrayrulewidth \futurelet
   \reserved@a\@xthickhline}
\def\@xthickhline{\ifx\reserved@a\thickhline
               \vskip\doublerulesep
               \vskip-\thickarrayrulewidth
             \fi
      \ifnum0=`{\fi}}
\makeatother

\newlength{\thickarrayrulewidth}
\usepackage{booktabs,makecell,multirow,tabularx}
\newcolumntype{L}{>{\RaggedRight\arraybackslash}X}
\newcolumntype{C}{>{\Centering\arraybackslash}X}

\usepackage{ragged2e}
\usepackage{amssymb}

\usepackage{tikz,xcolor,hyperref}

\definecolor{lime}{HTML}{A6CE39}
\DeclareRobustCommand{\orcidicon}{
    \begin{tikzpicture}
    \draw[lime, fill=lime] (0,0) 
    circle [radius=0.16] 
    node[white] {{\fontfamily{qag}\selectfont \tiny ID}};
    \draw[white, fill=white] (-0.0625,0.095) 
    circle [radius=0.007];
    \end{tikzpicture}
    \hspace{-2mm}
}

\foreach \x in {A, ..., Z}{\expandafter\xdef\csname orcid\x\endcsname{\noexpand\href{https://orcid.org/\csname orcidauthor\x\endcsname}
            {\noexpand\orcidicon}}
}

\title{Gradient Variance Loss for Structure-Enhanced Image Super-Resolution}
\name{Lusine~Abrahamyan$^{1,2}$\orcidD{}, Anh~Minh~Truong$^{2,3}$\orcidC{}, Wilfried~Philips$^{2,3}$\orcidB{} and Nikos~Deligiannis$^{1,2}$\orcidA{} 
\thanks{E-mail: lusine.abrahamyan@vub.be (L. Abrahamyan), ndeligia@etrovub.be (N. Deligiannis), wilfried.philips@ugent.be (W. Philips), anhminh.truong@ugent.be (A. Truong). This work was supported by the Research Foundation–Flanders (FWO) Research under Project G093817N.}}
\address{$^1$ETRO Department, Vrije Universiteit Brussel (VUB), Pleinlaan 2, B-1050 Brussels, Belgium \\
$^2$imec, Kapeldreef 75, B-3001 Leuven, Belgium \\
$^3$TELIN-IPI, Ghent University, St-Pietersnieuwstraat 41, B-9000 Gent, Belgium}
\begin{document}
%
\maketitle
\begin{abstract}

Recent success in the field of single image super-resolution (SISR) is achieved by optimizing deep convolutional neural networks (CNNs) in the image space with the L1 or L2 loss. However, when trained with these loss functions, models usually fail to recover sharp edges present in the high-resolution (HR) images for the reason that the model tends to give a statistical average of potential HR solutions. During our research, we observe that gradient maps of images generated by the models trained with the L1 or L2 loss have significantly lower variance than the gradient maps of the original high-resolution images. In this work, we propose to alleviate the above issue by introducing a structure-enhancing loss function, coined Gradient Variance (GV) loss,  and generate textures with perceptual-pleasant details. Specifically, during the training of the model, we extract patches from the gradient maps of the target and generated output, calculate the variance of each patch and form variance maps for these two images. Further, we minimize the distance between the computed variance maps to enforce the model to produce high variance gradient maps that will lead to the generation of high-resolution images with sharper edges. Experimental results show that the GV loss can significantly improve both Structure Similarity (SSIM) and peak signal-to-noise ratio (PSNR) performance of existing image super-resolution (SR) deep learning models.

\end{abstract}

\begin{figure}[ht]
\begin{center}
   \includegraphics[width=1.0\linewidth]{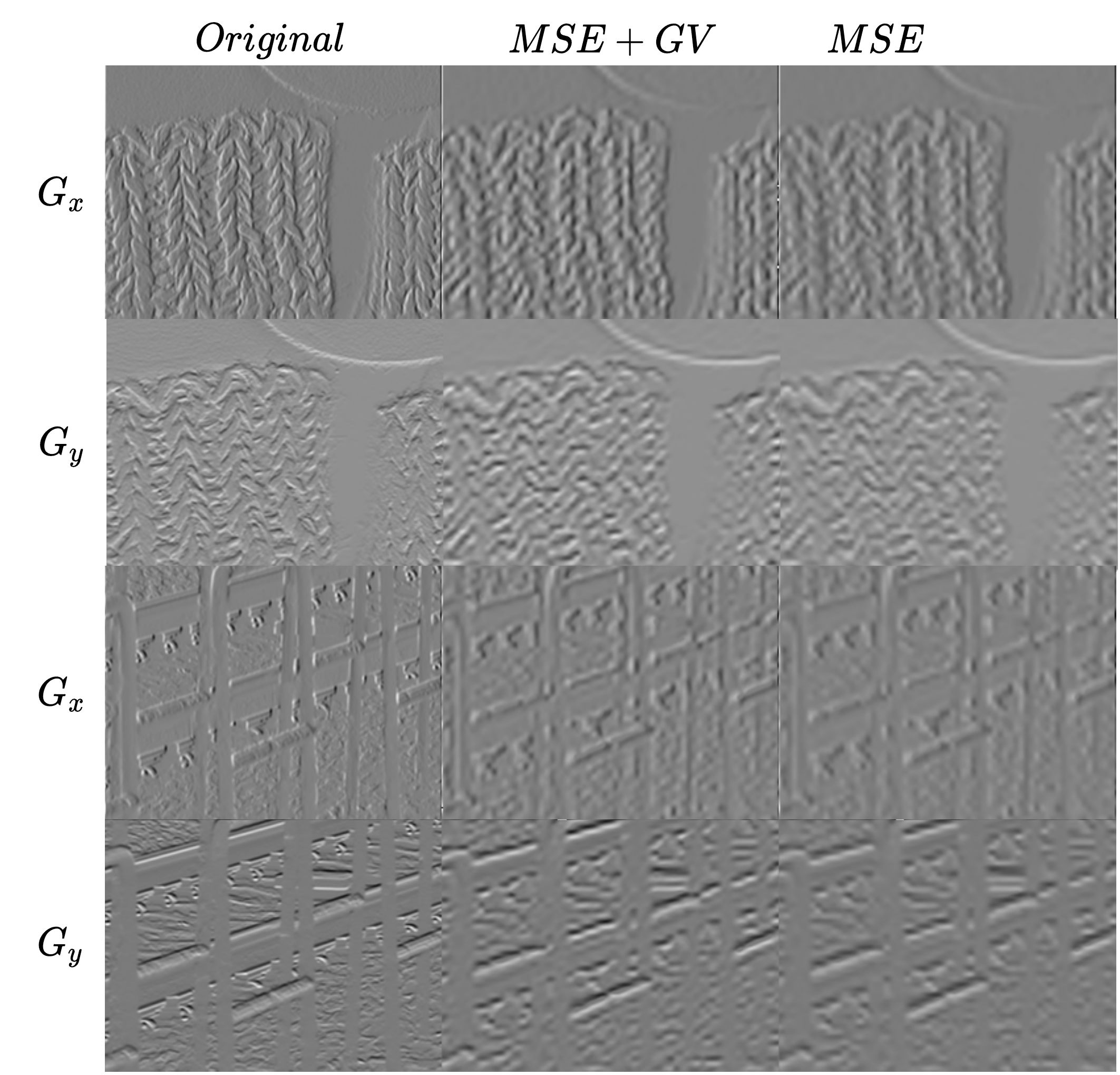}
\end{center}
   \caption{Gradient maps of sample images from DIV2K validation set for the VDSR model trained with only MSE loss and MSE$+$GV loss functions, where $G_{x}$ and $G_{y}$ represent the gradient maps on direction $X$ and $Y$, correspondingly. }
\label{fig:gradient_comparison}
\end{figure}

\section{Introduction}
\label{sec:intro}

Single image super-resolution (SISR) aims at producing high-resolution (HR) images from their low-resolution (LR) counterparts. It is a widely known ill-posed problem since each LR image may have many HR interpolations. With the advance of deep learning, a number of super-resolution (SR) methods based on convolutional neural networks (CNNs) have been proposed~\cite{shi2016realespcn, lim2017enhancededsr, wang2018esrgan, ma2020structure}. 
Many of them are optimized using the mean squared error (MSE) as the objective function, which aims at minimizing the pixel-wise distance between the predicted SR image and the corresponding HR image. Images estimated from the models optimized using the MSE as objective function can have high peak signal-to-noise ratios but often can lack high-frequency details and be perceptually unsatisfying and blurry.
Several objective functions based on generative adversarial networks (GANs)~\cite{wang2018esrgan, ledig2017photosrgan} have been proposed for SISR to generate photo-realistic images. In this case, the discriminator learns to differentiate between the generated and original images, while the generator tries to produce outputs on which the discriminator will fail to provide the correct prediction (i.e., fake or real). Nevertheless, these GAN based methods can lead to the generation of redundant and semantically inaccurate textures.

To overcome the issue of inaccurate texture generation, several researchers~\cite{ma2020structure, sun2008imageprior, sharpprior, zhu2015modeling} proposed utilizing the information contained in the gradient maps of the LR and HR images to generate structurally coherent outputs. In these works, gradients are used to construct the dictionaries~\cite{zhu2015modeling}, priors~\cite{sharpprior} or used as ground-truth data to training the part of the model responsible for learning the gradient representation of the HR image~\cite{ma2020structure}.
While these methods enhance the edges of the SR image using the gradient maps, they introduce new learning parameters and increase the latency of the models. 

In this work, we introduce a new loss function that utilizes gradient maps to enhance the edges of SR images. Specifically, we propose to minimize the distance between the variances of gradient maps obtained from the HR and SR images. Our experimental results indicate that the proposed gradient variance loss improves notably the performance of  existing models for SISR such as VDSR~\cite{kim2016accuratevdsr} and EDSR~\cite{lim2017enhancededsr}. 

The rest of the paper is organized as follows. 
Section~\ref{sec:RelatedWork} reviews the related work and Section~\ref{sec:gv_loss} presents the proposed gradient variance loss. Experimental results are presented in Section~\ref{sec:experiments} and conclusions are drawn in Section~\ref{sec:conclusion}.

\section{Related Work}
\label{sec:RelatedWork}
\subsection{Objective Functions for Image Reconstruction}
To ensure spatial continuity in the generated images and avoid noisy results, Gatys et al.~\cite{gatys2016image} proposed to use the total-variation (TV) loss in the optimization process. The TV loss is a regularization loss that tries to minimize the pixel-wise distance between the neighbouring pixels. While the TV loss helps to eliminate the overly pixelated results, at the same time, it promotes the generation of oversmoothed images that lack details.

Objective functions based on GANs also play an important role in the SR problem.
Ledig et al.~\cite{ledig2017photosrgan} adopts the adversarial loss~\cite{goodfellow2014generative} in the  SRGAN~\cite{ledig2017photosrgan} model to generate photo-realistic HR images. Furthermore, Sajjadi et al.~\cite{sajjadi2017enhancenet} proposed to restore high-fidelity textures by using the texture loss. While GAN based objectives functions boost the photo-realism of the generated images, they also introduce redundant textures and geometric distortions.

\subsection{Gradient Maps in the SISR task}
Several works proposed to utilize the gradient maps inferred from RGB images to guide image recovery. The authors of~\cite{ma2020structure} propose to add a separate branch to learn the gradient map of the HR image from the LR representation and fuse the information with the main branch. The research presented in~\cite{sun2008imageprior} used a gradient profile prior, which is a parametric prior describing the shape and the sharpness of the image gradients, to provide a constraint on image gradients when estimating an HR image. Yan et al.~\cite{sharpprior} proposed an SR model based on the gradient profile sharpness which is extracted from gradient description models. Furthermore, Zhu et al.~\cite{zhu2015modeling} proposed a gradient-based SR method by constructing a dictionary of gradient patterns and modeling deformable gradient compositions. 
While all these methods explore the gradient maps to improve the visual representation of the generated SR images, they add learnable parameters related to the gradient information in the model, which entails an increase in model complexity and  latency. 

Unlike the existing approaches, we propose to utilize the gradient information only in the optimization process. Specifically, we design the gradient-variance loss that uses the gradient maps of the SR and HR images to enhance the structure of the generated texture.

\section{Gradient Variance Loss}
\label{sec:gv_loss}

The goal of SISR is to estimate the high-resolution image $I^{SR}$ (a.k.a., the super-resolved image predicted by a model) given an interpolated low-resolution $I^{LR}$. In our setup, to produce $I^{LR}$ we downscale its high-resolution $I^{HR}$ counterpart by a factor of $s$ using bicubic interpolation. We refer to $s$ as the upscaling ratio. In general, both $I^{LR}$ and $I^{HR}$ can have $c$ colour channels, and for the $I^{LR}$ with the height $h$ and width $w$, they can be represented as tensors of size $c \times h \times w$ and $c\times s \cdot h \times s \cdot w$, respectively. 

In the recent years, the most successful approaches of obtaining the $I^{SR}$ image from the $I^{LR}$ image are the ones based on CNNs. In these CNN models, the $I^{LR}$ image is passed through a number of convolutional layers and upscaled using learnable layers (e.g. transposed convolution~\cite{gatys2016image}) or a signal-processing operations operations (e.g. pixel-shuffle~\cite{shi2016realespcn}, bicubic upsample).

Our empirical analysis reveals that the gradient maps of $I^{SR}$ images, predicted by CNN models,  and those of $I^{HR}$ have significant differences. Specifically, the gradient maps of $I^{HR}$ have sharper edges and are more detailed compared with the ones computed from $I^{SR}$ (see Figure~\ref{fig:gradient_comparison}). This behaviour can be seen for the models optimized in the image space by using the L2 loss alone. The reason is that minimizing the L2 loss encourages finding pixel-wise averages of plausible solutions, which can be overly smooth and hence have low perceptual quality~\cite{johnson2016perceptual, mathieu2015deep}.

In this work, we suggest using the variance of the gradient maps as a metric for enhancing the structure of the generated  $I^{SR}$ images during the optimisation process. 
Given $I^{SR}$ and $I^{HR}$, the corresponding gradient maps $G^{SR}_{x}$, $G^{SR}_{y}$, $G^{HR}_{x}$, $G^{HR}_{y}$ are calculated by converting the images from the RGB to the gray space and applying Sobel kernels on the grayscale representations. These gradient maps are then unfolded into $n \times n$ non-overlapping patches and form $\Tilde{G}^{SR}_{x}$, $\Tilde{G}^{SR}_{y}$, $\Tilde{G}^{HR}_{x}$, $\Tilde{G}^{HR}_{y}$ matrices with dimensions $n^2 \times \frac{w \cdot h}{n^{2}}$, where each column represents one patch. The $i$-th element of the variance maps of the matrices is calculated as 
\begin{equation}
    v_{i} = \Big(\frac{\sum_{j=1}^{n^2}(\Tilde{G}_{i,j} - \mu_{i})^{2}}{n^2 - 1}\Big), \text{  }i =1, \dots, \frac{w \cdot h}{n^2}
,
\end{equation}
where $\mu_{i}$ is the mean value of the $i$-th patch and $\Tilde{G}$ represents one of the unfolded gradient maps. Given the variance maps $v_{x}^{SR}, v_{y}^{SR}$ and $v_{x}^{HR}, v_{y}^{HR}$ for the corresponding  $I^{SR}$ and $I^{HR}$ images, the gradient variance loss can be formulated as
\begin{align}
    & L_{GV} = 
            \mathop{\mathbb{E}_{SR}} \|  v_{x}^{SR} - v_{x}^{HR} \|_2 + \mathop{\mathbb{E}_{SR}} \|  v_{y}^{SR} - v_{y}^{HR} \|_2.
\end{align}
The intuition behind the proposed loss function is the following:  gradient maps of generated images are blurry; hence, the variance of each individual region is lower than the variance of the same region on the $I^{HR}$ image. Therefore, when optimised with the proposed loss function, the model will be forced to provide sharper edges to minimise the difference between the variances of $I^{HR}$ and $I^{SR}$ images.

\begin{table*}[t]
\caption{Public benchmark test results and DIV2K validation results (PSNR(dB) / SSIM). Note that DIV2K validation results are acquired from published demo codes.}
\begin{center}
\scalebox{0.95}{
    \begin{tabular}{l||c||c|c||c|c||c|c }
    \thickhline
     Dataset    & Scale  & VDSR  & VDSR + GV & EDSR & EDSR + GV & ESPCN & ESPCN + GV \\
      \thickhline
              &  2$\times$  &27.41 / 0.8494&\textbf{28.00} / \textbf{0.8672}    &29.07 / 0.8847  &\textbf{29.40} / \textbf{0.8874}    &27.29 / 0.8667&\textbf{27.35} / \textbf{0.8695} \\
     Set14    &  3$\times$  &25.81 / 0.7866&\textbf{26.03} / \textbf{0.7943}    &22.45 / 0.6969  &\textbf{22.51} / \textbf{0.6987}    &24.45 / 0.7618&\textbf{24.54} / \textbf{0.7655}       \\
              &  4$\times$  &23.95 / 0.6993&\textbf{24.34} / \textbf{0.7181}    &\textbf{24.95} / 0.7346&24.93 / \textbf{0.7356} &23.87 / 0.7071&\textbf{23.95} / \textbf{0.7102}  \\
     \hline
              &  2$\times$ &30.38 / 0.9100&\textbf{31.21} / \textbf{0.9171}     &35.62 / 0.9599&\textbf{35.99} / \textbf{0.9614}    &30.42 / 0.9264&\textbf{30.55} / \textbf{0.9269} \\
     Set5     &  3$\times$ &28.67 / 0.8714&\textbf{28.94} / \textbf{0.8761}     &\textbf{26.49} / 0.8277 &26.48 / \textbf{0.8281}    &26.48 / 0.8456&\textbf{26.57} / \textbf{0.8469}  \\
              &  4$\times$ &26.22 / 0.8002&\textbf{26.77} / \textbf{0.8148}     &\textbf{30.47} / 0.8973&30.40 / \textbf{0.8987} &26.18 / 0.8075&\textbf{26.30} / \textbf{0.8078}  \\
     \hline
              &  2$\times$ &24.64 / 0.8148&\textbf{25.29} / \textbf{0.8361}     &25.98 / 0.8461  &\textbf{26.58} / \textbf{0.8505}     &24.73 / 0.8256&\textbf{24.75} / \textbf{0.8279} \\
     Urban    &  3$\times$ &23.18 / 0.7467&\textbf{23.41} / \textbf{0.7568}     &20.29 / 0.6734  & \textbf{20.32} / \textbf{0.6750}    &22.22 / 0.7129&\textbf{22.25} / \textbf{0.7169} \\
              &  4$\times$ &21.55 / 0.6525&\textbf{21.85} / \textbf{0.6721}     &\textbf{22.21} / 0.7167&22.14 / \textbf{0.7177}      &21.63 / 0.6548&\textbf{21.66} / \textbf{0.6575}\\
     \hline
              &  2$\times$ &30.13 / 0.8877 &\textbf{30.82} / \textbf{0.9036}   &34.24 / 0.9431   &\textbf{35.00} / \textbf{0.9485}    &30.32 / 0.9016  &\textbf{30.44} / \textbf{0.9043}\\
     DIV2K    &  3$\times$ &28.49 / 0.8439 &\textbf{28.58} / \textbf{0.8478}   &\textbf{31.29} / \textbf{0.8935}   &31.23 / 0.8929  &27.69 / 0.8294  &\textbf{27.79} / \textbf{0.8329} \\
    validation&  4$\times$ &26.47 / 0.7715 &\textbf{26.82} / \textbf{0.7853}   &29.29 / 0.8449&29.24 / \textbf{0.85456}   &26.42 / 0.7721  &\textbf{26.50} / \textbf{0.7746}  \\

      \thickhline
\end{tabular}}
\end{center}
\label{table:performance}
\end{table*}

\begin{figure}[t]   
\centering
\subfloat[]{\includegraphics[width=0.5\textwidth]{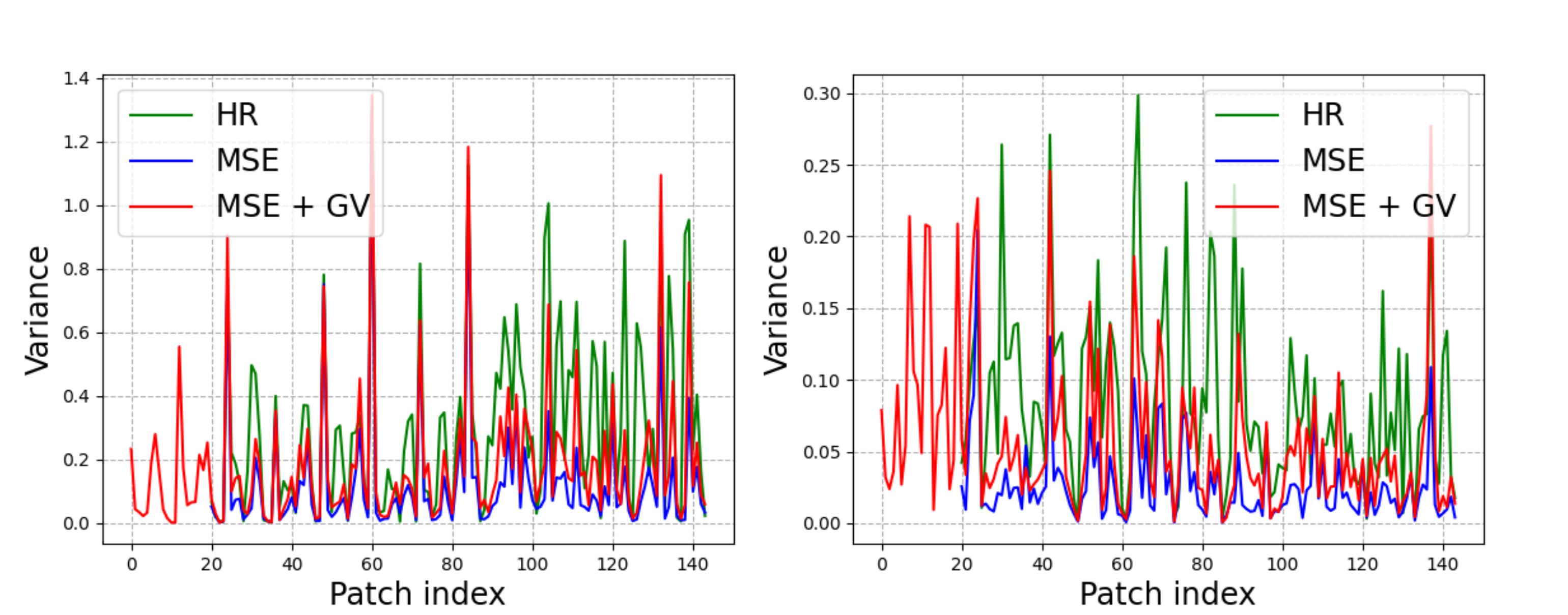}
\label{fig:plotx}}
\hfill
\subfloat[]{\includegraphics[width=0.5\textwidth]{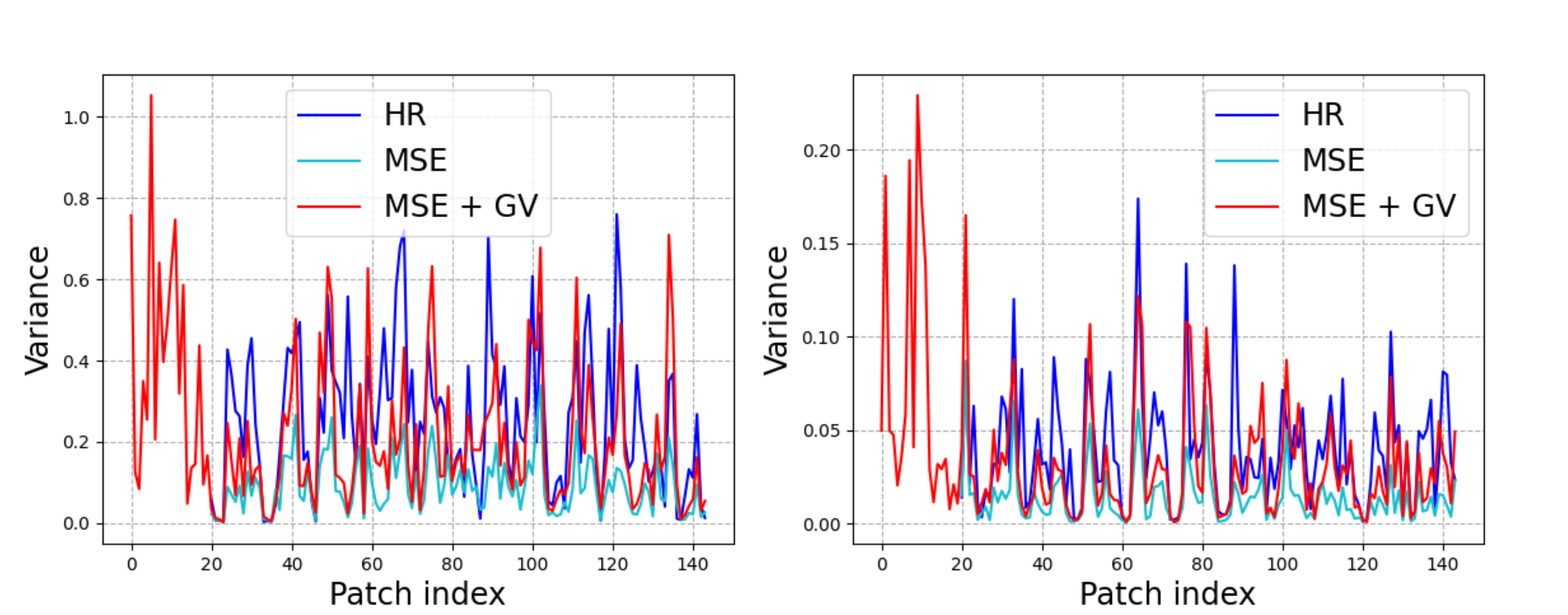}
\label{fig:ploty}}
\caption{Variances of $8\times8$ patches of the calculated gradient maps for images of the DIV2K validation set (a) for the gradients calculated in the X direction and (b) for the gradients calculated in the Y direction.}
\label{fig:plot_x}
\end{figure}

\section{Experiments}
\label{sec:experiments}

We present empirical results to demonstrate the efficiency of our GV loss. All experiments were performed on a single machine with 2 GeForce RTX 2080 Ti GPUs. We used PyTorch~\cite{pytorch} as a machine learning framework. Moreover, within all experiments, we used a fixed seed and only deterministic versions of the algorithms to ensure the reproducibility of the results. Further, for all the tested models, during the trainings, we used the combination of the GV loss with the objective function reported in the papers of the corresponding models.  

\subsection{Single Image Super Resolution}
\textbf{Datasets and Evaluation Metrics}: We evaluate the SR
performance of our proposed GV loss by utilizing
DIV2K~\cite{agustsson2017ntirediv2k} as the training dataset and three commonly used
benchmarks for testing: Set5~\cite{bevilacqua2012lowset5}, Set14~\cite{zeyde2010singleset14},
Urban100~\cite{huang2015singleurban100}. We downscale HR
images using bicubic interpolation to get LR inputs and
consider scaling factors of $2\times$, $3\times$ and $4\times$ in our experiments. We
choose the peak signal-to-noise ratio (PSNR) and structural similarity index measure (SSIM) as evaluation metrics. Higher PSNR and SSIM values indicate higher quality.

\textbf{Experiments and the Results}: We choose three well-performing SR models (EDSR~\cite{lim2017enhancededsr}, VDSR~\cite{kim2016accuratevdsr}, ESPCN~\cite{shi2016realespcn}) originally trained with the L2 and L1 loss functions, and perform additional trainings, with the hyperparameter setups described in each paper, where we additionally included the GV loss to enhance structural details in the reconstructed images. We use a patch size of $8 \times 8$ for the trainings with a scaling factor of $2\times$ and $16 \times 16$ for the trainings with $3\times$ and $4\times$ scaling factors. The results in Table~\ref{table:performance} indicate that for all tested models the proposed GV loss can boost the PSNR by $0.1-1.0$ dB. Specifically, for the $2\times$ scale factor on Set5, it increases the PSNR of the VDSR and EDSR models by $0.83$ dB and $0.67$ dB, respectively.

Moreover, we compare the GV loss with the total-variation (TV) loss function, which is widely used as a regularization in image reconstruction tasks. The results illustrated in Table~\ref{table:loss_trainings} show that our GV loss systematically outperforms the TV loss, when combined with the conventional loss functions used in the image reconstruction tasks.

\begin{table}
\caption{The PSNR and SSIM of VDSR model trained with the various loss functions.}
\begin{center}
\scalebox{0.9}{
    \begin{tabular}{l||c|c }
    \thickhline
     Loss                      & PSNR (dB)  & SSIM\\
      \thickhline
     L2 loss                    &  30.13        & 0.887736  \\
     L2 loss + TV loss          &  30.39      &  0.893212 \\
     L2 loss + GV loss          &  \textbf{30.82}     & \textbf{0.903685}  \\
     \hline
     L1 loss                    & 31.00       & 0.904594   \\
     L1 loss + TV loss          & 31.03      &  0.905105   \\
     L1 loss + GV loss          & \textbf{31.12}       & \textbf{0.907823}  \\
     \hline
     SSIM loss                    & 31.77        &  0.924859\\
     SSIM loss + TV loss          &  31.79      &  0.922874  \\
     SSIM loss + GV loss          &   \textbf{31.88}      &  \textbf{0.925011}\\

      \thickhline
\end{tabular}}
\end{center}
\label{table:loss_trainings}
\end{table}

\subsection{Analysis of the Variance}
To empirically evaluate the differences between the variances of gradient maps of SR and HR images, and show the advantage of using the GV loss, we performed the following experiment: for the ground-truth HR image, the image generated by the SR model trained with the $MSE$ loss, and the image generated by the model when trained with the $MSE+GV$ loss we calculate the variances of $8 \times 8$ patches of the gradient maps and compare them. From Figure~\ref{fig:plot_x}, it can be seen that: (1) the ground-truth HR images have higher peaks (i.e., higher values of variance) compared with the SR images generated using only the $MSE$ loss, which in general has smoothed plot, (2) the $MSE + GV$ loss setup can provide substantial boost to the variance of the generated SR image. 

\begin{figure}[ht!]
\begin{center}
  \includegraphics[width=1.0\linewidth]{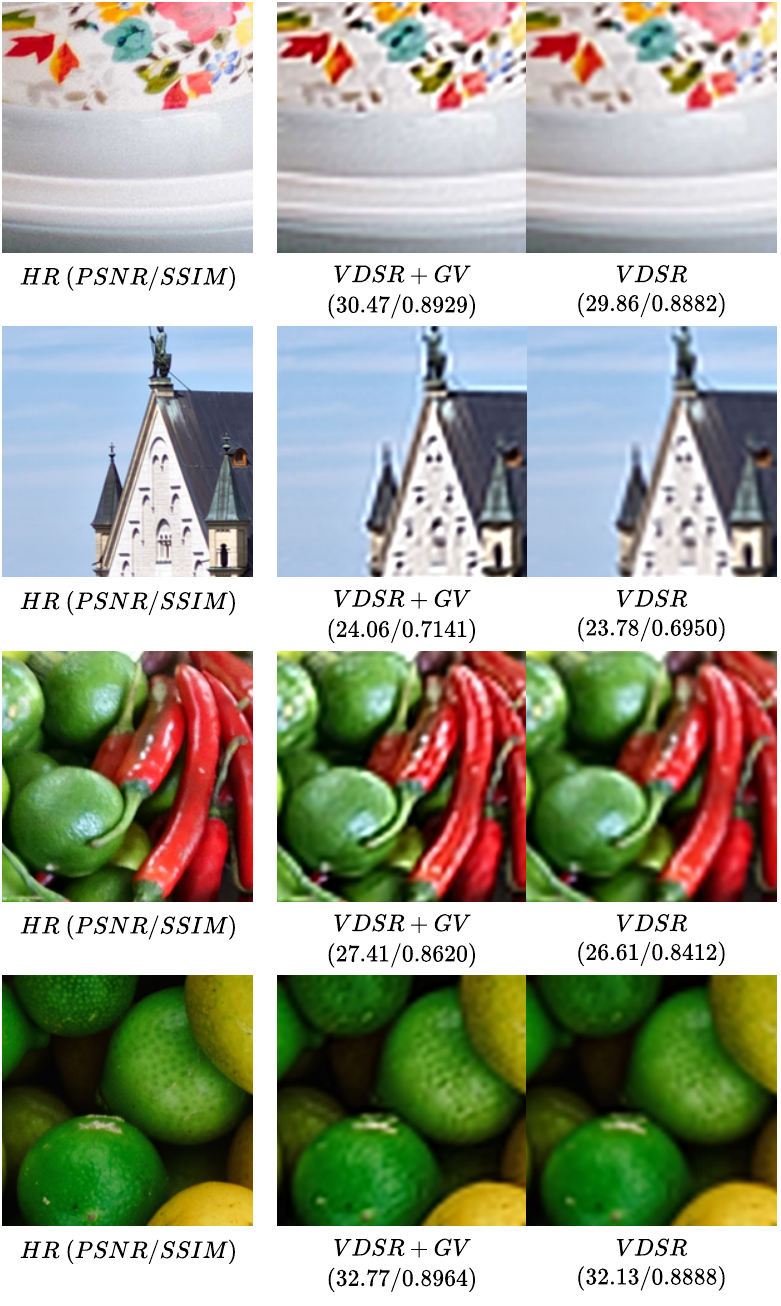}
\end{center}
  \caption{Comparison of the image patches from DIV2K validation set for the VDSR model trained with and without GV loss, with the upscaling ratio $s = 3$.}
\label{fig:img_comparison}
\end{figure}
    
\section{Conclusion}
\label{sec:conclusion}

In this paper, we have proposed a new structure-enhancing loss, coined gradient variance loss, for the SISR task to alleviate the issue of over-smoothness commonly existing in the generated SR images when trained with the L1 or L2 loss. Evaluation performed on the benchmark datasets, using well-performing super-resolution models, shows that the proposed loss function can systematically improve the widely used metrics of estimation the perceptual quality PSNR and SSIM.

\bibliographystyle{IEEEbib}
\bibliography{egbib}

\end{document}